\documentclass[
    ,final            % use final for the camera ready runs
  ]
  {aipproc}

\layoutstyle{6x9}

\newcommand{\ba}{\begin{eqnarray}}
\newcommand{\ea}{\end{eqnarray}}
\newcommand{\ban}{\begin{eqnarray*}}
\newcommand{\ean}{\end{eqnarray*}}
\newcommand{\bsub}{\begin{subequations}}
\newcommand{\esub}{\end{subequations}}

\newcommand{\nc}{\newcommand}
\nc{\Id}{{\mathchoice {\rm 1\mskip-4mu l} {\rm 1\mskip-4mu l}
{\rm 1\mskip-4.5mu l} {\rm 1\mskip-5mu l}}}

\begin{document}

\title{Supersymmetric structure in the Dirac~equation with
cylindrically-deformed~potentials}


\author{A. Leviatan}{
  address={
Racah Institute of Physics, The Hebrew University, 
Jerusalem 91904, Israel}
}

\begin{abstract}
Classes of relativistic symmetries accommodating supersymmetric 
patterns are considered for the Dirac Hamiltonian with axially-deformed 
scalar and vector potentials. 
\end{abstract}

\maketitle

%%%%%%%%%%%%%%%%%%%%%%%%%%%%%%%%%%%%%%%%%%%%
%% MAINMATTER
%%%%%%%%%%%%%%%%%%%%%%%%%%%%%%%%%%%%%%%%%%%%

The Dirac equation 
plays a key role in microscopic descriptions 
of many-fermion systems, employing 
covariant density functional theory and the 
relativistic mean-field approach. In application to nuclei 
and hadrons, the required Dirac mean-field Hamiltonian involves a mixture 
of Lorentz vector and scalar potentials. Recently, symmetries of 
Dirac Hamiltonians with such Lorentz structure were shown to be relevant 
for explaining observed degeneracies in the spectra of nuclei 
(pseudospin doublets~\cite{gino97}) and 
mesons (spin doublets~\cite{page01}). 
Corresponding supersymmetric patterns 
were identified for spherically-symmetric potentials~\cite{lev04}.
In the present contribution, we extend these studies to 
Dirac Hamiltonians with axially-deformed scalar and vector 
potentials~\cite{lev08}.
\begin{figure}[b]
  \includegraphics[height=.25\textheight,angle=-90]{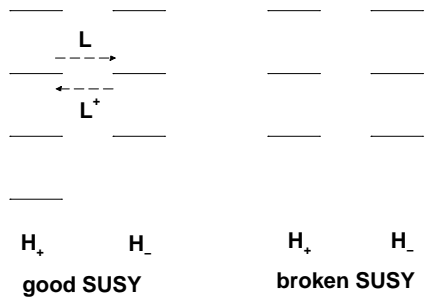}
  \caption{Typical spectra of good and broken SUSY. The operators $L$ and 
$L^{\dagger}$ connect degenerate states in the $H_{+}$ and $H_{-}$ 
sectors.}
\end{figure}
\begin{table}
\begin{tabular}{lll}
\hline
\noalign{\smallskip}
Class & \hspace{1.5cm}Symmetry & \hspace{1.5cm}Supersymmetry\\
\noalign{\smallskip}\hline\noalign{\smallskip}
(I) & $V_{S}(\rho,z) + V_{V}(\rho,z) = \Delta_0$ (pseudospin)
&
$\;\;$(good SUSY)\\
(II) & $V_{S}(\rho,z) - V_{V}(\rho,z) = \Xi_0$ (spin) &
$\;\;$(broken SUSY)\\
(III) & $V_S = V_{S}(z)$, $V_V = V_{V}(\rho)$ &
$\;\;V_S = V_{S}(z)$, $V_V = \frac{\alpha_{V}}{\rho}$ 
(good SUSY)\\
(IV) & $V_S = V_{S}(\rho)$, $V_V = V_{V}(z)$ &
$\;\;V_S = \frac{\alpha_{S}}{\rho}$, 
$\;\; V_V = V_{V}(z)$ (good SUSY)\\
\noalign{\smallskip}\hline
\end{tabular}\\
\caption{Classes of symmetries and related supersymmetries of the 
Dirac Hamiltonian with axially-deformed scalar, $V_{S}(\rho,z)$, and 
vector, $V_{V}(\rho,z)$, potentials~\cite{lev08}.}
\end{table}

The essential ingredients of supersymmetric quantum 
mechanics~\cite{witten81} are the supersymmetric Hamiltonian  
${\cal H} = \left ({H_{+}\quad\atop 0}{0\atop H_{-}}\right ) = 
\left ({L^{\dagger}L\atop 0}{0\atop LL^{\dagger}}\right )$ 
and 
charges 
$Q_{-} = \left ({ 0\atop L}{ 0\atop 0}\right )$, 
$Q_{+} = (Q_{-})^{\dagger}$, 
which generate the supersymmetric algebra 
\ba
\left [\,{\cal H}\,,\,Q_{\pm}\,\right ] = 
\left \{\,Q_{\pm}\,,\,Q_{\pm}\,\right \}=0\;\; , \;\; 
\left \{\,Q_{-}\,,\,Q_{+}\,\right \}= {\cal H} ~. 
\label{susy}
\ea 
$H_{+}$ and $H_{-}$ satisfy an 
intertwining relation, $LH_{+} = H_{-}L$, which ensures that 
if $\phi^{+}$ is an eigenstate of $H_{+}$, then also 
$\phi^{-}=L\phi^{+}$ is an eigenstate of $H_{-}$ with the 
same energy. Such pairwise degeneracy persists, 
unless $L\phi^{+}$ vanishes or produces an unphysical state, 
({\it e.g.}, non-normalizable), as occurs 
for the ground state when the supersymmetry (SUSY) is exact. 
The SUSY partner Hamiltonians 
$H_{+}$ and $H_{-}$ are therefore isospectral in the sense that 
their spectra consist of pairwise degenerate levels 
and a possible non-degenerate single state in one sector, when 
the supersymmetry is exact. If all states are pairwise degenerate, 
the supersymmetry is said to be broken. Typical spectra 
for good and broken SUSY are shown in Fig.~1.

The Dirac Hamiltonian, $H$, for  a fermion of mass M moving in external scalar 
$V_S$ and vector $V_V$ potentials is given by 
$H = {\bf \hat{\alpha}\cdot p}
+ \hat{\beta} (M  + V_S) + V_V$, 
where $\hat{\alpha}_i$, 
$\hat{\beta}$ are the usual Dirac matrices,  
and we have set 
$\hbar = c =1$. 
When the potentials are axially-symmetric, {\it i.e.}, independent of the 
azimuthal angle $\phi$, $V_{S,V}=V_{S,V}(\rho,z)\,$, 
$\rho = \sqrt{x^2+y^2}$, 
then the $z$-component of the angular momentum 
operator, $\hat{J}_z$, commutes with the Hamiltonian, 
and its eigenvalues $\Omega$ are used to label the Dirac wave functions
\ba
\Psi_{\Omega}(\rho,\phi,z) = 
\left ( 
\begin{array}{c}
g^{+}(\rho,z)\, e^{i(\Omega - 1/2)\phi}\\
g^{-}(\rho,z)\, e^{i(\Omega + 1/2)\phi}\\
if^{+}(\rho,z)\, e^{i(\Omega - 1/2)\phi}\\
if^{-}(\rho,z)\, e^{i(\Omega + 1/2)\phi}\\
\end{array}
\right ) ~.
\label{wf1}
\ea
Here $g^{\pm}(\rho,z)$ and $f^{\pm}(\rho,z)$ are radial wave functions 
for the upper and lower components respectively.
The Dirac equation, $H\Psi=E\Psi$, leads to four coupled partial 
differential equations for these functions. 
Their solutions are greatly simplified in the presence of symmetries. 
In what follows we discuss four classes of relativistic symmetries, 
listed in Table 1, and possible supersymmetries within each 
class~\cite{lev08}. 

The symmetry of class I, referred to as pseudospin symmetry, 
occurs when the sum of the scalar and vector 
potentials is a constant, 
$V_{S}(\rho,z) + V_{V}(\rho,z) = \Delta_0$. 
The symmetry generators, ${\hat{\tilde {S}}}_{i}$, 
commute with the Dirac Hamiltonian 
 and span an SU(2) algebra~\cite{bell75,ginolev98}
\ba
{\hat{\tilde {S}}}_{i} = 
\left (
\begin{array}{cc}
 U_p\, \hat{s}_i\, U_p &  0 \\
0 & \hat{s}_{i}
\end{array}
\right ) 
\quad i =x,y,z 
\qquad U_p = \, {\vec{\sigma}\cdot \vec{p} \over p} ~.
\label{pSgen}
\ea
Here 
${\hat s}_{i} = \sigma_{i}/2$ are the usual spin operators, defined 
in terms of Pauli matrices. 
The Dirac eigenfunctions in the pseudospin limit satisfy
\ba
{\hat{\tilde {S}}}_{z}
\Psi^{(\tilde{\mu})}_{\Omega} &=&
\tilde{\mu}\Psi^{(\tilde{\mu})}_{\Omega}\qquad
\;\tilde{\mu} = \pm 1/2
\ea
and form degenerate $SU(2)$ doublets. 
\begin{figure}
\includegraphics[width=\linewidth]{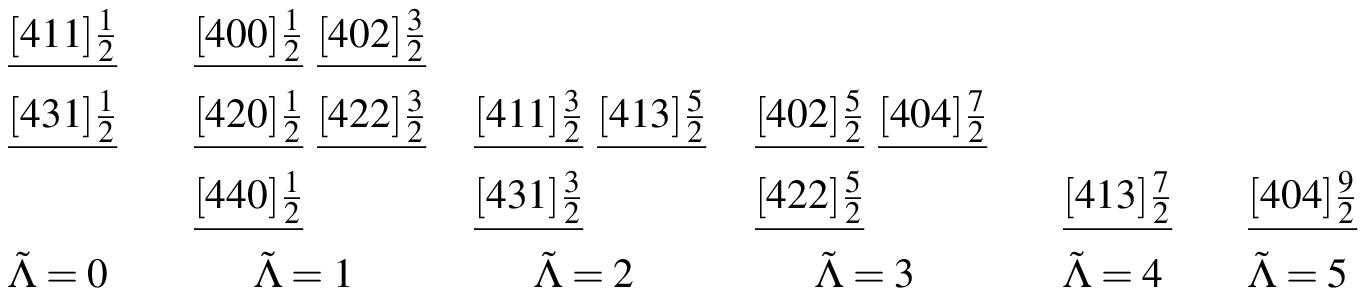} 
\caption{Grouping of deformed shell-model states
$[N=4,n_3,\Lambda]\Omega$, exhibiting a pattern of good SUSY relevant 
to the pseudospin symmetry limit. 
$N$ and $n_3$ are harmonic oscillator quantum numbers. 
$\tilde{\Lambda}$ 
is the pseudo-orbital angular momentum projection along the symmetry 
$z$-axis.}
\end{figure}
Their wave functions have the form~\cite{gino02}
\ba
\Psi^{(-1/2)}_{\Omega_1=\tilde{\Lambda}-1/2} 
= 
\left ( 
\begin{array}{c}
g^{+}\, e^{i(\tilde{\Lambda} - 1)\phi}\\
-g\, e^{i\tilde{\Lambda}\phi}\\
0 \\
if\, e^{i\tilde{\Lambda}\phi}\\
\end{array}
\right )
\quad , \quad
\Psi^{(1/2)}_{\Omega_2=\tilde{\Lambda}+1/2} 
= 
\left ( 
\begin{array}{c}
g\, e^{i\tilde{\Lambda}\phi}\\
g^{-}\, e^{i(\tilde{\Lambda}+1)\phi}\\
if\, e^{i\tilde{\Lambda}\phi}\\
0 \\
\end{array}
\right )
\label{wfps}
\ea
where $\tilde{\Lambda} = \Omega-\tilde{\mu}$ 
is the eigenvalue of $\hat{J}_z - {\hat{\tilde {S}}}_{z}$.
The relativistic pseudospin symmetry has 
been tested in numerous realistic mean field calculations 
of nuclei and were found to be obeyed to a good approximation, 
especially for doublets near the Fermi surface~\cite{ginlev04,gino05}. 
The dominant upper components of the states in Eq.~(\ref{wfps}),  
involving $g^{+}$ and $g^{-}$, correspond to non-relativistic 
pseudospin doublets with asymptotic (Nilsson) quantum numbers 
$[N,n_3,\Lambda]\Omega=\Lambda+1/2$ and
$[N,n_3,\Lambda +2]\Omega=\Lambda+3/2$, respectively. 
The doublet is expressed in terms of the pseudo-orbital angular momentum 
projection, $\tilde{\Lambda}=\Lambda+1$, 
which is added to the pseudospin projection, $\tilde{\mu}=\pm 1/2$, 
to form doublet states with  $\Omega=\tilde{\Lambda}\pm 1/2$. 
Such doublets play a crucial role in explaining features of 
deformed nuclei, including superdeformation and identical 
bands~\cite{gino05}. The doublet structure can be deduced 
from the fact that the present Dirac Hamiltonian commutes 
with ${\hat{\tilde {S}}}_{z}$ and $L$, 
\ba
L &=& 
2\left (\,M+\Delta_0 - H\, \right )\,
{\hat{\tilde {S}}}_{x} ~,
\label{Lpsx}
\ea
while $L$ and ${\hat{\tilde {S}}}_{z}$ anticommute.
The operator $L$ connects the two states in Eq.~(\ref{wfps}) 
and $L^2 = 
L^{\dagger}L = 
(\, M+\Delta_0 - H\,)^2$. 
Consequently, near the pseudospin limit, 
the spectrum, for each $\tilde{\Lambda}\neq 0$, 
consists of twin towers of pairwise degenerate pseudospin doublet states,  
with $\Omega_1=\tilde{\Lambda}-1/2$ and $\Omega_2=\tilde{\Lambda}+1/2$, 
and an additional non-degenerate 
nodeless state at the bottom of the $\Omega_1=\tilde{\Lambda}-1/2$ tower, 
whose pseudospin partner state is not a bound Dirac eigenstate~\cite{lev08}. 
Altogether, the ensemble of Dirac states with $\Omega_2-\Omega_1=1$ exhibits 
a supersymmetric pattern of good SUSY, as illustrated schematically 
in Fig.~2. 

The symmetry of class II, referred to as spin symmetry, occurs 
when the difference of the scalar and vector potentials is a constant,  
$V_{S}(\rho,z) - V_{V}(\rho,z) = \Xi_0$. 
The discussion of spin symmetry is similar to that of pseudospin symmetry 
with the role of upper and lower components interchanged. Specifically, 
the symmetry group is again $SU(2)$ and its generators~\cite{bell75}
\ba
{\hat{S}}_{i} = 
\left (
\begin{array}{cc}
\hat{s}_{i} & 0 \\
0 & U_p\, \hat{s}_i\, U_p
\end{array}
\right ) 
\quad i =x,y,z ~
\label{Spgen}
\ea
commute with the Dirac Hamiltonian. 
The Dirac eigenfunctions in the spin limit satisfy
\ba
\hat{S}_{z}
\Psi^{(\mu)}_{\Omega} &=&
\mu\,\Psi^{(\mu)}_{\Omega}\qquad
\;\mu = \pm 1/2
\ea
and form degenerate $SU(2)$ doublets. Their wave functions 
are of the form~\cite{gino05}
\ba
\hspace{-1cm}
\Psi^{(1/2)}_{\Omega_{1}=\Lambda+1/2} 
= 
\left ( 
\begin{array}{c}
g\, e^{i\Lambda\phi}\\
0 \\
if\, e^{i\Lambda\phi}\\
if^{-}\, e^{i(\Lambda+1)\phi}\\
\end{array}
\right )
\quad , \quad
\Psi^{(-1/2)}_{\Omega_{2}=\Lambda-1/2} 
= 
\left ( 
\begin{array}{c}
0 \\
g\, e^{i\Lambda\phi}\\
if^{+}\, e^{i(\Lambda - 1)\phi}\\
-if\, e^{i\Lambda\phi}
\end{array}
\right )
\label{wfsp}
\ea
where $\Lambda = \Omega -\mu\geq 0$ is the eigenvalue of 
$\hat{J}_z - \hat{S}_{z}$.
The upper components of the two states in Eq.~(\ref{wfsp}) 
form the usual non-relativistic spin doublet with a common radial wave 
function~$g$, an orbital angular momentum projection, 
$\Lambda$, and two spin orientations 
$\Omega = \Lambda\pm 1/2$.
The above relativistic spin symmetry is relevant to the structure of 
heavy-light quark mesons~\cite{page01}. 
Its characteristic doublet structure can be deduced 
from the fact that the present Dirac Hamiltonian commutes with 
the operators $\hat{S}_{z}$ (\ref{Spgen}) and $L$ 
\ba
L &=& 2\left (\,M+\Xi_0 + H\, \right )\,\hat{S}_{x} ~, 
\label{Lspx}
\ea
while $L$ and $\hat{S}_{z}$ anticommute.
\begin{figure}
  \includegraphics[width=\textwidth]{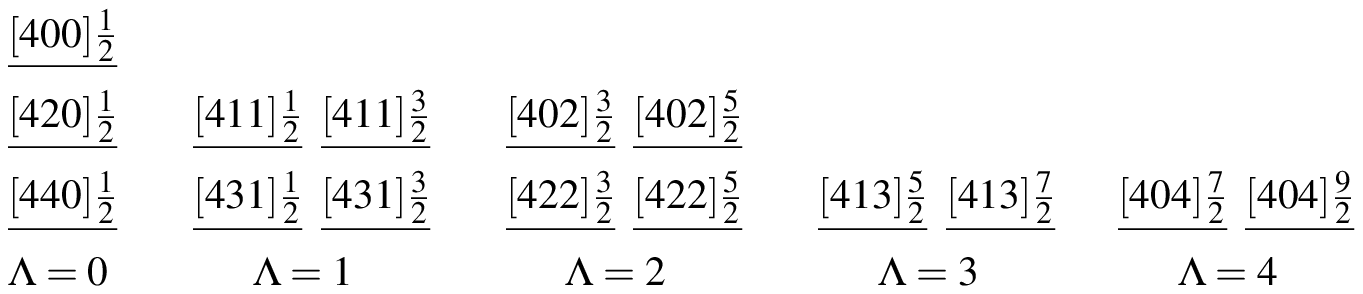}
\caption{Grouping of deformed shell-model states
$[N=4,n_3,\Lambda]\Omega$, exhibiting a pattern of broken SUSY relevant 
to the spin symmetry limit. $\Lambda$ is the orbital 
angular momentum projection.}
\end{figure}
The operator $L$ 
connects the two states in Eq.~(\ref{wfsp}) 
and $L^2 = 
L^{\dagger}L = 
(\, M+\Xi_0 + H\,)^2$. 
These relations imply that near the spin limit, 
the spectrum, for each $\Lambda\neq 0$ 
consists of twin towers of pairwise degenerate spin-doublet states 
with $\Omega_1=\Lambda-1/2$ and $\Omega_2=\Lambda+1/2$. 
None of these towers have a 
single non-degenerate state. 
Altogether, the ensemble of Dirac states with 
$\Omega_2-\Omega_1=-1$ exhibits a supersymmetric pattern 
which resembles that of a broken SUSY, as illustrated in Fig.~3. 

The Dirac Hamiltonian has additional symmetries when the scalar and 
vector potentials depend on different variables. 
The symmetry of class III occurs when  
the potentials are of the form
$V_{S}=V_{S}(z)$ and $V_{V}=V_{V}(\rho)$. 
In this case, the Dirac Hamiltonian commutes with the following 
Hermitian operator 
\ba
R_{z} &=& \left [\,M+V_{S}(z)\,\right ]\hat{\beta}\,\hat{\Sigma}_3 
-i\gamma_{5}\frac{\partial}{\partial z} ~, 
\label{Rz}
\ea 
where $\hat{\Sigma}_i = \left ({\sigma_i\atop 0}{0\atop \sigma_i}\right )$. 
The Dirac eigenfunctions satisfy
\ba
R_z\,\Psi^{(\epsilon)} &=& \epsilon\,\Psi^{(\epsilon)} ~.
\label{RzPsi}
\ea
A separation of variables 
is possible by choosing the Dirac 
wave function in the form 
\ba
\Psi^{(\epsilon)} 
&=& \frac{1}{\sqrt{\rho}} 
\left ( 
\begin{array}{c}
u_{1}(\rho)v_{1}(z)\,
e^{i(\Omega -1/2)\phi}\\
u_{2}(\rho)v_{2}(z)\,
e^{i(\Omega +1/2)\phi}\\
iu_{1}(\rho)v_{2}(z)\,
e^{i(\Omega -1/2)\phi}\\
-iu_{2}(\rho)v_{1}(z)\,
e^{i(\Omega +1/2)\phi}
\end{array}
\right ) ~.
\label{PsiRz}
\ea
\noindent
The Dirac equation then reduces to a set of two coupled first-order 
ordinary differential equations in the variable $\rho$, 
and a separate set in the variable $z$. 
The wave functions depend parametrically on the quantum number 
$\epsilon$ (\ref{RzPsi}), which can be continuous or discrete, and 
$\vert\epsilon\vert$ plays 
the role of a mass for the transverse motion.  
A particular selection of potentials within the 
symmetry class III, 
is relevant to the study of electron channeling 
in crystals ($V_{S}(z) = 0$) \cite{greiner94}. 
A supersymmetric pattern within the present class is obtained 
for the choice
$V_{V}(\rho) = \alpha_{V}/\rho$ and $V_{S}(z)$ arbitrary. 
The energy eigenvalues are 
$E^{(\epsilon)}_{n_{\rho},\Omega} = 
|\epsilon|/\sqrt{
1 + \alpha_{V}^2/(n_{\rho}+\gamma)^2}\;$ 
$(n_{\rho} =0,1,2,\ldots)$, 
with $\gamma = \sqrt{\Omega^2 - \alpha_{V}^2}$.
For $n_{\rho}\geq 1$, the states 
$\Psi^{(\pm\epsilon)}_{n_{\rho},\Omega}$ are degenerate. 
For $n_{\rho}=0$ only one state is an acceptable solution, 
which has $\epsilon>0$ 
(assuming $\alpha_V < 0$). For each $\Omega$ and $\epsilon$ the spectrum 
resembles a supersymmetric pattern of good SUSY, with the towers $H_{+}$ 
($H_{-}$) of Fig.~1 corresponding to states with 
$\epsilon>0$ ($\epsilon<0$).  
The explicit solvability and observed degeneracies in the present 
case can be attributed to the existence of an additional Hermitian 
operator 
\ba
L &=& \hat{\beta}\,\hat{\Sigma}_3
\left \{
i\hat{J}_{z}\gamma_5\,
\left [\,H -  \hat{\Sigma}_3R_z\, \right ]
-\frac{\alpha_{V}}{\rho}\,(\vec{\Sigma}\cdot\vec{\rho}\,)\,
R_z\;  \right \}
\label{LRz}
\ea
which commutes with $H$ but anticommutes with $R_z$ (\ref{Rz}).  
The operator $L$ connects the doublet states with 
$(n_{\rho}\geq1,\pm\epsilon)$ 
and annihilates the single state with ($n_{\rho}=0,\epsilon>0)$. 
In addition, 
$L^2 = L^{\dagger}L = 
\hat{J}_{z}^2 (\, H^2 - R_{z}^2\,) + \alpha_{V}^2\,R_{z}^2$.

The symmetry of class IV occurs when the potentials are of the form 
$V_{S}=V_{S}(\rho)$ and $V_{V}=V_{V}(z)$. 
Its discussion is similar to that of class III, with the $\rho$- and 
$z$-dependence of the potentials interchanged. 
In this case, the following Hermitian operator 
\ba
R_{\rho} &=&
[\,M+V_{S}(\rho)\,]\hat{\Sigma}_3 
-i\hat{\beta}\,\gamma_{5}
(\,\vec{\hat{\Sigma}}\times\vec{\hat{p}}\,)_{3}
\label{Rrho}
\ea
commutes with the Dirac Hamiltonian and the 
Dirac eigenfunctions satisfy
\ba
R_{\rho}\,\Psi^{(\tilde{\epsilon})} &=& 
\tilde{\epsilon}\,\Psi^{(\tilde{\epsilon})} ~.
\label{RrhoPsi}
\ea
Again, a separation of variables is possible with the choice of 
wave function, $\Psi^{(\tilde{\epsilon})}$, in a form similar to 
that of Eq.~(\ref{PsiRz}). 
The quantum number $\tilde{\epsilon}$ 
plays now the role of an energy for the transverse motion. 
A particular selection of potentials within the symmetry class IV 
was encountered in the study of the Schwinger mechanism for 
particle-production in a strong confined 
field  ($V_{V}(z) =\alpha_{V}z$) \cite{wang88}, 
$q\bar{q}$ pair-creation in a flux tube 
($V_{S}(\rho=0)$ \cite{pavel91}, and the 
canonical quantization in cylindrical geometry of a free Dirac field 
($V_{S}(\rho)=V_{V}(z)=0$) \cite{cooper06}. 
A supersymmetric pattern within the present class is obtained for 
the choice $V_{S}(\rho) = \alpha_{S}/\rho$ $(\alpha_{S} <0)$ 
and $V_{V}(z)$ arbitrary.   
The allowed values are
$\tilde{\epsilon} = 
\pm M\sqrt{
1 -\alpha_{S}^2/(n_{\rho}+\tilde{\gamma})^2}$
$(n_{\rho}=0,1,2.\ldots)$, 
where $\tilde{\gamma} = \sqrt{\Omega^2 + \alpha_{S}^2}$. 
For $n_{\rho}\geq 1$ the eigensolutions with $\pm\tilde{\epsilon}$ 
are degenerate, $E^{(\tilde{\epsilon})}_{n_{\rho},\Omega} = 
E^{(-\tilde{\epsilon})}_{n_{\rho},\Omega}$. 
For $n_{\rho}=0$ only one state, with $\tilde{\epsilon}>0$,  
is an acceptable solution. 
For each $\Omega$ and $\tilde{\epsilon}$ the resulting spectrum 
resembles a supersymmetric pattern of good SUSY. As before, the 
indicated structure is due to an additional Hermitian operator 
\ba
L &=& 
\hat{\Sigma}_3
\left \{
i\hat{J}_{z}\gamma_5\,
\left [\,M - \hat{\Sigma}_3 R_{\rho}\,\right ]
-\frac{\alpha_{S}}{\rho}
(\vec{\Sigma}\cdot\vec{\rho}\,)
\hat{\beta}R_{\rho}\;  \right \}
\label{LRrho}
\ea
which commutes with $H$ but anticommutes with $R_{\rho}$ (\ref{Rrho}). 
The operator 
$L$ connects the doublet states with $(n_{\rho}\geq 1,\pm\tilde{\epsilon})$ 
and annihilates the single state with ($n_{\rho}=0,\tilde{\epsilon}>0)$. 
In addition, 
$L^2 = L^{\dagger}L = 
\hat{J}_{z}^2 (\, R_{\rho}^2 - M^2\,)
+ \alpha_{S}^2\,R_{\rho}^2$.

The supersymmetric patterns obtained within each of the four classes of 
symmetries share a common origin, namely, the existence of two 
conserved and anticommuting operators $R$ and $L$. 
Here $R=\hat{\tilde {S}}_{z},\,\hat{S}_{z},\,R_{z},\,R_{\rho}$, 
given in Eqs.~(\ref{pSgen}),(\ref{Spgen}),(\ref{Rz}),(\ref{Rrho}) 
and $L$ is given in 
Eqs.~(\ref{Lpsx}),(\ref{Lspx}),(\ref{LRz}),(\ref{LRrho}), 
for classes (I)-(IV) respectively. 
The operator $R$ has eigenvalues which come in 
pairs of opposite signs.
Denoting the eigenvalue of $R$ by $r$, we can define an 
operator ${\cal P}_r= R/\vert r\vert $ 
with eigenvalues $\pm 1$, which provide a ``parity'' label to 
eigenstates of the corresponding Dirac Hamiltonian, $H$. 
The operators $L$, ${\cal P}_{r}$ and $H$ satisfy 
$\left [\, H\, , \, {\cal P}_r\,\right ] =
\left [\, H\, , \, L\,\right ] = 
\left \{\, {\cal P}_r\, , \, L\,\right \} =0$. 
The operator $L$ connects eigenstates of opposite ``parity'' 
and its square, $L^2=L^{\dagger}L=f(H)$, 
is a quadratic function of $H$. 
Supersymmetric charges ($Q_{\pm}$) 
and Hamiltonian (${\cal H}$), which satisfy 
the supersymmetric algebra (\ref{susy}),   
can now be constructed as 
$Q_{\pm} = \frac{1}{2}L\left (1 \mp {\cal P}_r \right )$ 
and ${\cal H} = L^2 = f(H)$.
The operator ${\cal P}_r$ serves as a $Z_2$-grading operator 
accompanying the supersymmetric algebra. 
This work is supported by the Israel Science Foundation.

\end{document}